\documentclass[12pt]{article}
\usepackage{amsmath,amsfonts,amssymb,latexsym,graphicx,caption}
\DeclareGraphicsExtensions{.pdf,.jpg,.PNG,.JPG}

\numberwithin{equation}{section}
\def\be{\begin{equation}}
\def\ee{\end{equation}}
\def\bq{\begin{eqnarray}}
\def\eq{\end{eqnarray}}
\def\beq{\begin{eqnarray}}
\def\eeq{\end{eqnarray}}

\def\a{\alpha}
\def\b{\beta}

\def\pa{\partial}

\begin{document}

\title{\Large{\textsc{Regular braneworlds with nonlinear bulk-fluids}}}
\author{{\large\textsc{Ignatios Antoniadis$^{1,2}$\thanks{\texttt{antoniad@lpthe.jussieu.fr}}
Spiros Cotsakis$^{3,4}$\thanks{\texttt{skot@aegean.gr}}
Ifigeneia Klaoudatou$^{4}$\thanks{\texttt{iklaoud@aegean.gr}}}} \\
$^1$Laboratoire de Physique Th\'{e}orique et Hautes Energies - LPTHE\\
Sorbonne Universit\'{e}, CNRS 4 Place Jussieu, 75005 Paris, France\\
$^2$ Institute for Theoretical Physics, KU Leuven\\ Celestijnenlaan 200D, B-3001 Leuven, Belgium\\
$^{3}$Institute of Gravitation and Cosmology, RUDN University\\
ul. Miklukho-Maklaya 6, Moscow 117198, Russia\\
$^{4}$Research Laboratory of Geometry,  Dynamical Systems  and Cosmology\\
University of the Aegean, Karlovassi 83200, Samos, Greece}
\maketitle
\begin{abstract}
\noindent
We construct a regular five-dimensional brane-world with localised gravity on a flat 3-brane. The matter content in the bulk is parametrised by an analog of a non-linear fluid with equation of state $p=\gamma\rho^\lambda$
between the `pressure' $p$ and the `density' $\rho$ dependent on the 5th dimension. For $\gamma$ negative and $\lambda>1$, the null energy condition is satisfied and the geometry is free of singularities
within finite distance from the brane, while the induced four-dimensional Planck mass is finite.
\end{abstract}
\newpage
\tableofcontents
\newpage
\section{Introduction}
In previous works~\cite{ack1}-\cite{ack5}, we have made a detailed investigation
of the singularity structure of a five-dimensional (5d) brane-world, consisting on
a 3-brane embedded in five dimensions with bulk matter content parametrised
by an analog of a perfect fluid with equation of state $p=\gamma\rho$,
between the `pressure' $p$ and the `density' $\rho$ dependent on the 5th
coordinate $Y$, playing the role of time. This work extended previous studies of
brane-worlds with bulk scalar fields~\cite{nima, silverstein, gubser}. The parameter $\gamma$ should be
greater than $-1$, in order to satisfy the null energy condition (guaranteeing
the absence of ghosts in the bulk). It turns out that for a flat brane, there is
always a singularity that appears in the bulk within finite distance from the
brane position~\cite{ack2,ack3}. The singularity can be avoided by considering
half of the space, starting from the brane position to the side which is
singularity free and using appropriate junction (matching) conditions~\cite{ack3}. The resulting construction however does not localise gravity on the brane and the effective 4d Planck mass is divergent.
These results essentially continue to hold when the brane is curved and
maximally symmetric. In this case, the null energy condition restricts
$\gamma$ to be in the interval $(-1,-1/2)$ and the brane is negatively
curved (anti-de Sitter), but still does not localise gravity~\cite{ack5}.

In this work, we extend our analysis by considering a flat brane and replacing
the bulk content with an analog of a non-linear fluid, satisfying an equation
of state $p=\gamma\rho^\lambda$. Formally similar equations of state  have been used in various cosmological contexts such as  in studies of escaping big-rip singularities
during late time asymptotics \cite{srivastava, diaz, nojiri}, in obtaining inflationary models with special properties \cite{ba90}, as unified models of dark energy and dark matter \cite{ka, sen}, as well as in studies of singularities \cite{ck1, ck2, not}.

We show that for $\gamma$ negative and $\lambda>1$, the null energy
condition is satisfied and there is no singularity within finite distance from
the brane. Moreover, we are able to find explicit general solutions in terms of
hypergeometric functions that further simplify considerably for particular
values of $\lambda=1+1/(2k)$ with $k$ a positive integer. It is important to
emphasise that these solutions are regular independently of the presence of
the brane. When the latter is taken into account a cutting and matching
procedure is needed for half of the space, as explained above, which can be
done in a way that the integral for the induced 4d effective Planck mass is
finite, leading to localised gravity on the brane for any value of the exponent
$\lambda>1$. In the perfect fluid limit $\lambda\to 1$, the 4d Planck mass
diverges and in agreement with the previous analysis.

The structure of this paper is as follows. In the next Section, we write down the system of dynamical equations holding in the bulk ($Y$-coordinate), together with the nonlinear equation of state assumed for the bulk fluid. This equation introduces a new parameter $\lambda$ into the problem that plays a central role in the following. We also  derive the precise  restrictions imposed by the validity of the energy conditions. In Section 3, we introduce the desirable properties of the solutions and show how these depend crucially on $\lambda$. In the remaining Sections 4-7, we study in detail the properties of solutions which have both the sought for properties, namely, they are free of finite $Y$-distance singularities and have a finite Planck mass. We construct these solutions and also show how their matching is done to produce the said behaviour. We conclude our work in the last Section.
\section{Evolution equations and consistency}
In this Section, we setup the field equations of a brane embedded in a 5-bulk
space which is filled with an imperfect fluid with a nonlinear equation of state,
and write down the restrictions imposed by the energy conditions on the bulk
fluid quantities by the 5-dimensional geometry.
\subsection{Field equations}
Our braneworld model consists of a 3-brane embedded in a five-dimensional
bulk space with a metric of the form,
\be
\label{warpmetric}
g_{5}=a^{2}(Y)g_{4}+dY^{2},
\ee
where $a(Y)$ is the warp factor with $a(Y)>0$, which we simply denote by $a$,
while $g_{4}$ is the four-dimensional flat metric,
{\it i.e.},
\be
\label{branemetrics}
g_{4}=-dt^{2}+dx_1^2+dx_2^2+dx_3^2.
\ee
Our sign conventions are defined as follows. We take capital Latin indices
$A,B,\dots=1,2,3,4,5$, Greek indices $\a,\b,\dots =1,2,3,4$,  with $t$ being
the timelike coordinate, and $(x_1,x_2,x_3,Y)$ the remaining spacelike ones.
The 5-dimensional Riemann tensor is defined by the formula,
\be
R^{A}_{\,\,\,BCD}=\pa_{C}\Gamma^{A}_{\,\,\,BD}-\pa_{D}\Gamma^{A}_{\,\,\,BC}+\Gamma^{M}_{BD}\Gamma^{A}_{MC}-\Gamma^{M}_{BC}\Gamma^{A}_{MD}
\ee
the Ricci tensor is the contraction,
\be
R_{AB}=R^{C}_{\,\,\,ACB},
\ee
and the five-dimensional Einstein equations on the bulk space are given by,
\be
G_{AB}=R_{AB}-\frac{1}{2}g_{AB}R=\kappa^{2}_{5}T_{AB}.
\ee
We assume a bulk fluid having an energy-momentum tensor of the form,
\be
\label{T old}
T_{AB}=(\rho+p)u_{A}u_{B}-p g_{AB},
\ee
with a nonlinear equation of state,
\be
\label{eos}
p=\gamma \rho^{\lambda},
\ee
where the `pressure' $p$ and the `density' $\rho$ are functions only of the
fifth coordinate $Y$ and $\gamma$, $\lambda$ are constants. The fluid velocity vector field is $u_{A}=(0,0,0,0,1)$, that is $u_{A}=\pa/\pa Y$, parallel to the $Y$-dimension.

The Einstein equations 
and the equation of energy-momentum conservation,
$
\nabla_{A}T^{AB}=0$, then become,
\bq
\label{syst2i}
\frac{a'^{2}}{a^{2}}&=&\frac{\kappa_{5}^{2}}{6}\rho,\\
\label{syst2ii}
\frac{a''}{a}&=&-\frac{\kappa_{5}^{2}}{6}{(2\gamma\rho^{\lambda}+\rho)},\\
\label{syst2iii}
\rho'&+&4(\gamma\rho^{\lambda}+\rho)\frac{a'}{a}=0.
\eq
where the prime $(')$ denotes differentiation with respect to $Y$.
Solving the system of equations Eq.~(\ref{syst2i}), Eq.~(\ref{syst2ii}) and Eq.~(\ref{syst2iii}),
entails constructing a scheme of solutions based on the ranges of the parameters $\gamma$
and $\lambda$. Our goal is to determine the ranges of parameters that lead to solutions that satisfy
physical conditions such as the energy conditions. To this end,
we first study the energy conditions and the restrictions that they imply on the density and pressure
of the fluid, we then express them in terms of the ranges of the parameters and finally we include them
in our solutions. In this way, we integrate from the beginning the energy conditions in our final forms of
solutions.
\subsection{Energy conditions}
We now move on to study the restrictions imposed by the strong, weak and null energy conditions on the bulk fluid (\ref{T old}). These translate to conditions to be satisfied by $p$ and $\rho$, and constitute \emph{a priori} restrictions on our solutions in the form of physical conditions.

In our formulation of the field equations, the metric (\ref{warpmetric}) and the bulk fluid appear as static with respect to the time coordinate $t$, because the evolution is taken with respect to the fifth spatial coordinate $Y$. We have however the unitarity constraint imposed on the model with regard to the $t$-evolution under which our 5d fluid is anisotropic in the $Y$-direction. Here, we take into account this positive energy constraint on the brane-bulk system following the analysis of~\cite{ack3}.


To work out the possible restrictions imposed on the system by the requirement
of consistency of the evolution in the $\pa/\pa t$ direction with that along the
$\pa/\pa Y$ vector field, we shall use the energy conditions. To proceed, we
reinterpret our fluid analogue as a real anisotropic fluid having the following energy momentum tensor,
\be
\label{T new}
T_{AB}= (\rho^{0}+p^{0})u_{A}^{0}u_{B}^{0}-p^{0}g_{\alpha\beta}\delta_{A}^{\alpha}\delta_{B}^{\beta}-
p_{Y}g_{55}\delta_{A}^{5}\delta_{B}^{5},
\ee
where $u_{A}^{0}=(a(Y),0,0,0,0)$, $A,B=1,2,3,4,5$ and $\alpha,\beta=1,2,3,4$.
When we combine (\ref{T old}) with (\ref{T new}), we get the following set of relations,
\bq
\label{p y to rho}
p_{Y}&=&-\rho\\
\label{rho new}
\rho^{0}&=&-p\\
\label{p new}
p^{0}&=&p.
\eq
The last two relations imply that
\be
\label{p new to rho new}
p^{0}=-\rho^{0},
\ee
which means that this type of matter satisfies a cosmological constant-like equation of state.
Substituting (\ref{p y to rho})-(\ref{p new to rho new})
in (\ref{T new}), we find that
\be
T_{AB}= -p g_{\alpha\beta}\delta_{A}^{\alpha}\delta_{B}^{\beta}+
\rho g_{55}\delta_{A}^{5}\delta_{B}^{5}.
\ee
We are now ready to form the energy conditions for our type of matter.
We begin with the weak energy condition according to which, every future-directed
timelike vector $v^{A}$ should satisfy
\be
T_{AB}v^{A}v^{B}\geq 0.
\ee
This condition implies that the energy density should be non negative for all forms of
physical matter \cite{wald}. Here we find (see \cite{ack3} and \cite{ack5}
for detailed proofs) that it translates to
\be
\label{wec_p}
p\geq 0, \quad \textrm{or, equivalently,}\quad  \gamma\geq 0
\ee
and
\be
\label{wec_1}
p+\rho\geq 0.
\ee
Inputting the equation of state $p=\gamma\rho^{\lambda}$ in Eq.~(\ref{wec_1}) we find
\be
\gamma \rho^{\lambda}+\rho=\rho^{\lambda}(\gamma+\rho^{1-\lambda})\geq 0
\ee
and since $\rho\geq 0$ from Eq.~(\ref{syst2i}), we simply get
\be
\gamma+\rho^{1-\lambda}\geq 0.
\ee

The strong energy condition, on the other hand, demands that
\be
\left(T_{AB}-\dfrac{1}{3}T g_{AB}\right)v^{A}v^{B}\geq 0,
\ee
for every future-directed unit timelike vector $v^{A}$. In our case,
this condition leads to the following restrictions for $p$ and $\rho$,
\be
\label{sec_1}
-p+\rho\geq 0,
\ee
and
\be
\label{sec_2}
p+\rho\geq 0.
\ee
Again, Eq.~(\ref{sec_1}) can be written equivalently
\be
-\gamma \rho^{\lambda}+\rho=\rho^{\lambda}(-\gamma+\rho^{1-\lambda})\geq 0
\ee
and using the fact that $\rho\geq 0$, we get
\be
-\gamma+\rho^{1-\lambda}\geq 0.
\ee
Similarly, from Eq.~(\ref{sec_2}) we find
\be
\gamma+\rho^{1-\lambda}\geq 0.
\ee

Finally, according to the null energy condition, every future-directed
null vector $k^{A}$ should satisfy \cite{poisson}
\be
T_{AB}k^{A}k^{B}\geq 0.
\ee
Here we find that it translates to
\be
p+\rho\geq 0,
\ee
which implies, as we have already seen, that
\be
\label{nec}
\gamma+\rho^{1-\lambda}\geq 0.
\ee
This equation together with the Friedman equation (\ref{syst2i}) give the following bound for the scale factor as a function of the $Y$ coordinate,
\be
a(Y)\geq e^{\sqrt{\rho_{\textrm{min}}}\, Y},
\ee
where $\rho_{\textrm{min}}=(-\gamma)^{1/(1-\lambda)}$.

To compare between the effects of a linear and non-linear equation of state
on the restrictions set by the energy conditions, we find it useful to mention here that for the case
of a linear equation of state ($\lambda=1$), the weak energy condition
requires $p\geq0$ and $\gamma>0$, while the strong energy condition restricts $p$
and $\gamma$ to either $p<0$ and $-1\leq\gamma<0$, or, $p>0$ and $0<\gamma\leq 1$ \cite{ack3}.

Summarizing the constraints set by the energy conditions for a non-linear fluid,
we see that the restriction described by (\ref{nec}) is required by all energy conditions
and we shall take this into account in the next Section in the procedure of deriving our solutions.
\section{Representation of solutions depending on  $\lambda$}
In this  Section, we give general properties of the solutions of equations Eq.~(\ref{syst2i}), Eq.~(\ref{syst2ii}) and Eq.~(\ref{syst2iii}).
The desirable properties we seek for these solutions are:
\begin{itemize}
\item regularity
\item localization of gravity on the brane.
\end{itemize}
For regularity, solutions should not have finite-$Y$-distance singularities, whereas the localization property is effected with a finite Planck mass. We shall discuss these two properties at length in later Sections.

\subsection{Some properties of the  $\lambda$-solutions}
Since the continuity equation (\ref{syst2iii}) is a separable equation, by integrating it and substituting into the Friedman equation (\ref{syst2i}), we can arrive at a general  relation between the density
and the warp factor satisfied by the system (\ref{syst2i}), (\ref{syst2ii}) and (\ref{syst2iii}).

While integrating, a logarithmic term of the form $\ln |\gamma+\rho^{1-\lambda}|$ appears.
We ignore the absolute value and simply
put this term equal to $\ln (\gamma+\rho^{1-\lambda})$, in compliance with the  restriction (\ref{nec}) of the previous Section as required by all energy conditions.
We the end up with,
\be
\label{rho to a}
\rho={(-\gamma+c_{1}a^{4(\lambda-1)})}^{1/(1-\lambda)},
\ee
where
\be c_1=\frac{\gamma+\rho_0^{1-\lambda}}{a_0^{4(\lambda-1)}},
\ee
with $\rho_0=\rho(Y_0), a_0=a(Y_0)$  the initial conditions.
We note that (\ref{nec}) now implies that $c_{1}\geq 0$, while
the strong energy condition translates to,
\be
\label{strong energy cond}
-2\gamma+c_{1}a^{4(\lambda-1)}\geq 0.
\ee
It is important to note from Eq.~(\ref{rho to a}) that $\rho$ becomes infinite for $\lambda>1$,
and
\be\label{sing}
a^{4(\lambda-1)}=\dfrac{\gamma}{c_{1}}.
\ee
To avoid this singularity, we are going to focus only on negative values of $\gamma$, which means
that the weak energy condition is not going to be satisfied by our solutions. However, the null energy
condition is already embodied in our solutions and the strong energy condition (\ref{strong energy cond})
is automatically satisfied for $\gamma<0$.

We continue by substituting Eq.~(\ref{rho to a}) in Eq.~(\ref{syst2i}), to obtain,
\be
\label{hyper_integral}
\int\dfrac{a}{(c_{1}-\gamma a^{4(1-\lambda)})^{1/(2(1-\lambda))}}\,da=\pm\dfrac{\kappa_{5}}{\sqrt{6}}\int dY.
\ee
Further examination of this equation reveals that solutions depend crucially on
whether $\lambda$ is greater or smaller than one. All  solutions can be grouped
in these two large classes, and this classification is also consistent with the
dynamics around finite equilibria, or at infinity, cf. \cite{ackDS21}.

For $\lambda$  greater than one, the integral on the left-hand side of Eq. (\ref{hyper_integral}) can
be directly integrated out for those values of $\lambda$ that make $1/(2(1-\lambda))$ a negative
integer. This is possible when $\lambda=(n+1)/n$, with $n=2k$ and $k$ a positive integer.

Take for example the simplest case $n=2$ which gives $\lambda=3/2$.
Then $1/(2(1-\lambda))=-1$ and we can directly integrate Eq. (\ref{hyper_integral})
to get the solution in the following implicit form
\be
\label{sol_3/2}
\pm Y+c_{2}=\dfrac{\sqrt{6}}{\kappa_{5}}\left(\dfrac{c_{1}}{2} a^{2}-\gamma\ln a\right),
\ee
where $c_{2}$ is an integration constant.
We can now go one step further and derive the general form of solution for $\lambda=(n+1)/n$, $n=2k$ and $k$ a
positive integer. Using the binomial theorem, integrating and wrapping up
terms we find,
\be
\label{sol_endpoints}
\pm Y+c_{2}=\dfrac{\sqrt{6}}{\kappa_{5}}\left(\sum_{s=0}^{k-1}\dfrac{k!}{(k-s)!s!}\dfrac{c_{1}^{k-s}}{2-2s/k} a^{2-2s/k}(-\gamma)^{s}+(-\gamma)^{k}\ln a\right).
\ee

For all values of $\lambda$  such that $\lambda\neq (n+1)/n$, the integral on
the left hand-side of Eq.~(\ref{hyper_integral}) cannot in general be calculated
directly. In this case, we can express the solutions in terms of the Gaussian
hypergeometric function $_{2}F_{1}(\alpha,b,c;z)$, defined by
\be
_{2}F_{1}(\alpha,b,c;z)=\frac{\Gamma(c)}{\Gamma(\alpha)\Gamma(b)}
\sum_{n=0}^{\infty}\frac{\Gamma(\alpha+n)\Gamma(b+n)}{\Gamma(c+n)}\frac{z^{n}}{n!}
\ee
and convergent for $|z|<1$, where $\Gamma$ is the Gamma function given by 
\be
\Gamma(z)=\int^{\infty}_{0}e^{-t}t^{z-1}dt, \quad Re(z)>0.
\ee
An integral representation of $_{2}F_{1}(\alpha,b,c;z)$ is \cite{Erdelyi}
\be
\label{integral theory}
_{2}F_{1}(\alpha,b,c;z)=\frac{\Gamma(c)}{\Gamma(b)\Gamma(c-b)}\int_{0}^{1}
t^{b-1}(1-t)^{c-b-1}(1-tz)^{-\alpha}dt,
\ee
provided that
\be
\label{hyper cond1}
0<Re(b)<Re(c).
\ee
The RHS of (\ref{integral theory}) is a one-valued analytic function $z$ within the domain
\be
\label{argument(1-z)}
|\arg(1-z)|<\pi.
\ee
This converges for $|z|<1$. However, because of the definition of $z$ and the
fact that on the circle of convergence there is a singular point given
by Eq. (\ref{sing}), it follows that by analytic continuation we can extend
the disk of convergence far beyond this limit (cf. eg., \cite{mf}, pp. 389-90).

To derive our solution in terms of a hypergeometric function,
we simply set
\be
z=\gamma a^{4(1-\lambda)}/c_{1},
\ee
and note that the restriction (\ref{argument(1-z)})
translates into,
\be
|\arg(1-z)|=\left|\arg\left(\dfrac{c_{1}-\gamma a^{4(1-\lambda)}}{c_{1}}\right)\right|<\pi,
\ee
which is satisfied since $-\gamma$ and $c_{1}$ are both positive real numbers. We arrive at the following form of our solutions,
\beq
&&\pm Y+c_{2}=\frac{\sqrt{6}}{4(1-\lambda)\kappa_{5}}c_{1}^{1/(2(\lambda-1))}a^{2}\int_{0}^{1}w^{1/(2(1-\lambda))-1}
\left(1-\frac{\gamma}{c_{1}}a^{4(1-\lambda)}w\right)^{-1/(2(1-\lambda))}dw\nonumber\\
&=&\frac{\sqrt{6}}{2\kappa_{5}}c_{1}^{1/(2(\lambda-1))}a^{2}\,_{2}F_{1}\left(\dfrac{1}{2(1-\lambda)},\dfrac{1}{2(1-\lambda)},
\dfrac{1}{2(1-\lambda)}+1;
\dfrac{\gamma}{c_{1}}a^{4(1-\lambda)}\right), \quad \lambda<1\label{a_sing},
\eeq
where $c_{2}$ is an integration constant. The restriction $\lambda<1$ above, follows from the condition
for a valid integral representation of $_{2}F_{1}$ described by
(\ref{hyper cond1}). This means that in order to derive solutions for
$\lambda>1$, we have to proceed in a different way (see Section 4).

In the next Subsection, we shall focus on the solutions having $\lambda<1$ which suffer from
a finite-distance singularity.
\subsection{Impossibility of viable solutions for $\lambda<1$}
It is straightforward to see that letting $a\rightarrow 0^{+}$ in Eq.~(\ref{a_sing})
leads to $Y\rightarrow \pm c_{2}$, which in turn implies that a collapse singularity appears within finite distance $c_{2}$.
On approach to the collapse singularity, it follows from Eq.~(\ref{rho to a})
that the density becomes divergent while Eq.~(\ref{eos}) implies then that the pressure
vanishes, or, diverges depending on whether $\lambda<0$, or, $0<\lambda<1$, respectively.

From our experience with similar models in previous work in \cite{ack5}, we find that
it is worth exploring the possibility of avoiding the collapse singularity
by finding out if it is the only type of finite-distance singularity
that solution Eq.~(\ref{a_sing}) can exhibit, or, if instead it coexists with a big-rip type of singularity
characterized by a divergent warp factor.

To examine the latter possibility, we study the behavior of the hypergeometric
function as $a\rightarrow \infty$, by making use of the following transformation
valid for $|\arg(-z)|<\pi$ (see \cite{Erdelyi}, pp. 63-64):
\be
\label{erdelyi}
_{2}F_{1}\left(\alpha,\alpha,\alpha+1;z\right)=-\alpha(-z)^{-\alpha}\,
\sum_{n=1}^{\infty}\dfrac{\Gamma(\alpha+n)}{n (n!)\Gamma(\alpha)}z^{-n}
+(-z)^{-\alpha}(\ln (-z)+\psi (1)-\psi (\alpha)),
\ee
where $\Gamma$ is the Gamma function and $\psi$ is its logarithmic derivative  \cite{Erdelyi},
\be
\psi(z)=\dfrac{d}{dz}\ln\Gamma(z)=\dfrac{\Gamma'(z)}{\Gamma (z)}.
\ee
Note that the hypergeometric function in (\ref{erdelyi}) becomes a real valued function for those real numbers $z$
that satisfy the requirement $|\arg(-z)|<\pi$ that we mention above, thus for all
non-positive real numbers $z$. The transformation of the hypergeometric function of (\ref{a_sing}) according
to (\ref{erdelyi}) gives,
\be\label{hyper-infty-lambda<1}
_{2}F_{1}\left(\dfrac{1}{2(1-\lambda)},\dfrac{1}{2(1-\lambda)},\dfrac{1}{2(1-\lambda)}+1;
\dfrac{\gamma}{c_{1}}a^{4(1-\lambda)}\right)=
\left(\dfrac{-\gamma}{c_{1}}\right)^{1/(2(\lambda-1))}a^{-2}\times \phi (a),
\ee
with,
\beq
\phi(a)&=&\dfrac{1}{2(\lambda-1)}\sum_{n=1}^{\infty}
\dfrac{\Gamma(1/(2(1-\lambda))+n)}{\Gamma(1/(2(1-\lambda)))n(n!)}
\left(\dfrac{\gamma}{c_{1}}a^{4(1-\lambda)}\right)^{-n}+\ln\left(-\dfrac{\gamma}{c_{1}}a^{4(1-\lambda)}\right)\nonumber\\
&+&\psi(1)-\psi\left(\dfrac{1}{2(1-\lambda)}\right).
\eeq
Substituting (\ref{hyper-infty-lambda<1}) in (\ref{a_sing}) we find
\beq
\nonumber
\pm Y+c_{2}&=&\dfrac{\sqrt{6}(-\gamma)^{1/(2(\lambda-1))}}{2\kappa_{5}}\left(\dfrac{1}{2(\lambda-1)}\sum_{n=1}^{\infty}
\dfrac{\Gamma(1/(2(1-\lambda))+n)}{\Gamma(1/(2(1-\lambda)))n(n!)}\left(\dfrac{\gamma}{c_{1}}a^{4(1-\lambda)}\right)^{-n}+\right. \\
&+&\left.\ln\left(-\dfrac{\gamma}{c_{1}}a^{4(1-\lambda)}\right)+\psi(1)-\psi\left(\dfrac{1}{2(1-\lambda)}\right)\right).
\eeq
It clearly follows from the above equation that letting $a\rightarrow\infty$
and since $\lambda<1$, the series appearing in the first term inside the bracket
converges to zero, but the logarithmic term that follows becomes divergent,
making the RHS approach infinity. Hence we conclude that $Y\rightarrow \pm \infty$.

This shows that the singularity with a
divergent warp factor shifts to infinite distance. On the other hand, the
behaviors of the density and pressure of the fluid following from
Eqs.~(\ref{rho to a}) and (\ref{eos}), are $\rho\rightarrow (-\gamma)^{1/(1-\lambda)}$
and $p\rightarrow -(-\gamma)^{1/(1-\lambda)}$.

The fact that for $\lambda<1$ there is only one type of finite-distance singularity, i.e a collapse singularity at $Y=\pm c_{2}$,
opens-up the possibility of using a matching mechanism, similar to the one we have
used for the case of a linear fluid in \cite{ack3} and \cite{ack5}, to construct a solution
free from finite-distance singularities for this range of $\lambda$.
However, the problem with localizing gravity on the brane that we faced in \cite{ack3},
persists here as well.

This  makes the choice $\lambda<1$ less promising exactly because
it leads to such a compromise. We therefore avoid giving details of the matching procedure for $\lambda<1$,
and focus instead on the more interesting available choice of $\lambda>1$.

We note  that the specific value $\lambda=1$ will also not be studied here, as it is
already analyzed in detail in our previous work \cite{ack1}-\cite{ack5}. There we showed that
a linear equation of state leads to power-law behavior for the warp factor.
Actually, one can also show that assuming a power-law behavior for the warp factor
and density and substituting these in Eq.~(\ref{syst2i}), (\ref{syst2ii}) and (\ref{syst2iii}),
forces $\lambda$ to be equal exactly equal to one.

In the next Sections, we are going to show that the case of solutions with $\lambda>1$,
not only offers the possibility of a regular solution but it also resolves the
problem of the localization of gravity that we faced in \cite{ack3} and \cite{ack5}.

\section{The $\lambda>1$ solutions}
In this Section, we give details about the construction and regularity properties of the solutions having  $\lambda>1$. This represents the most interesting case covered by our analysis.

\subsection{Construction}
For values of $\lambda$ greater than one, the requirement for the integral representation
previously met in the case $\lambda<1$, is no longer satisfied and we therefore have to elaborate further on the integral in
the LHS of Eq.~(\ref{hyper_integral}) to get a valid representation of a hypergeometric function.
This involves applying first integration by parts from which we find,
\beq
\nonumber
\label{int 1-3/2}
\int\dfrac{a}{(c_{1}-\gamma a^{4(1-\lambda)})^{1/(2(1-\lambda))}}\,da=
\dfrac{a^{2}}{2}(c_{1}-\gamma a^{4(1-\lambda)})^{1/(2(\lambda-1))}&-& \\
-\gamma\int a^{5-4\lambda}(c_{1}-\gamma a^{4(1-\lambda)})^{(3-2\lambda)/(2(\lambda-1))}\,da
\eeq
and then expressing the new integral on the RHS of the above equation as an integral representation
of a hypergeometric function, with a procedure similar to the one we did in Section 3.1, for $\lambda<1$.
While performing such a procedure and in order to satisfy restriction
(\ref{hyper cond1}), we have to confine the range of $\lambda$ to values
greater than $3/2$. We arrive at the following solution
\beq
\label{sol_3/2 above}
\nonumber
\pm Y+c_{2}&=&\dfrac{\sqrt{6}}{\kappa_{5}}\left(\dfrac{a^{2}}{2}(c_{1}-\gamma a^{4(1-\lambda)})^{1/(2(\lambda-1))}-
 \dfrac{\gamma c_{1}^{(3-2\lambda)/(2(\lambda-1))}}
{2(3-2\lambda)}a^{2(3-2\lambda)}\times \right. \\
& &\left.
_{2}F_{1}\left(\dfrac{3-2\lambda}{2(1-\lambda)},\dfrac{3-2\lambda}{2(1-\lambda)},\dfrac{3-2\lambda}{2(1-\lambda)}+1;\dfrac{\gamma}{c_{1}}a^{4(1-\lambda)}\right)\right),
\eeq
In (\ref{sol_3/2 above}), we have used the following integral representation
\be
\nonumber
_{2}F_{1}\left(\dfrac{3-2\lambda}{2(1-\lambda)},\dfrac{3-2\lambda}{2(1-\lambda)},\dfrac{3-2\lambda}{2(1-\lambda)}+1;\dfrac{\gamma}{c_{1}}a^{4(1-\lambda)}\right)=
\ee
\be
=\frac{2(1-\lambda)}{3-2\lambda}\int_{0}^{1}w^{(3-2\lambda)/(2(1-\lambda))-1}\left(1-\frac{\gamma}{c_{1}}a^{4(1-\lambda)} w\right)^{-(3-2\lambda)/(2(1-\lambda))}\,dw.
\ee
We note that for $1<\lambda<3/2$, the parameters of the hypergeometric function become
negative and restriction (\ref{hyper cond1}) is not satisfied. This means that
if we wish to find solutions for such values of $\lambda$, we cannot accept the integral
representation of the hypergeometric function as implied by Eq.~(\ref{sol_3/2 above}).

Finding a solution for $1<\lambda<3/2$,
becomes somewhat more intricate. The reason is, that in order to obtain a valid
representation of a hypergeometric function, we have to keep track with
condition (\ref{hyper cond1}) that requires the $\lambda$-dependent parameters
of the hypergeometric function, to be positive. This results in confining $\lambda$
to appropriate intervals, which in turn amounts to slicing the interval $(1,3/2)$ into
subintervals.

To proceed, we consider separately ranges of values of $\lambda$ such as $5/4<\lambda<3/2$, $7/6<\lambda<5/4$ and in general
$(n+1)/n<\lambda<(n-1)/(n-2)$, with $n=2k$ and $k$ a positive integer.
As we consider values of $\lambda$ closer and closer to one, we construct
our solution by taking the integral that was expressed as a representation
of a hypergeometric function in one interval, say $((n-1)/(n-2),(n-3)/(n-4))$,
and performing one more step of integration by parts to derive
a solution for its left-adjacent interval, that is $((n+1)/n,(n-1)/(n-2))$.

We end up with solutions expressed by hypergeometric functions of the
form $_{2}F_{1}(\alpha,\alpha,\alpha+1;z)$, where $\alpha=(n-n\lambda+1)/(2(1-\lambda))$,
and $z=(\gamma/c_{1})a^{4(1-\lambda)}$, and valid for $(n+1)/n<\lambda<(n-1)/(n-2)$.

We therefore see that as we move from one interval $((n-1)/(n-2),(n-3)/(n-4))$ to the next $((n-1)/(n-2),(n-3)/(n-4))$,
we obtain a solution expressed by a hypergeometric function that has a different
parameter $\alpha$ every time, but which maintains the same argument $z=(\gamma/c_{1})a^{4(1-\lambda)}$, with
$\lambda>1$.

Each of these hypergeometric series is convergent for $|z|<1$,
which in our case translates to $a^{4(\lambda-1)}>-\gamma/c_{1}$, where the bound
$-\gamma/c_{1}$ can of course be made arbitrarily large.

In the solutions we
obtained through this method, we shall  consider asymptotic behaviors
of the form $a\rightarrow \infty$, ensuring in this way convergence of
\emph{all} hypergeometric
series that appear, independently of the interval of validity of the solutions.

To see this clearly, consider for example the range $5/4<\lambda<3/2$. We
take the integral on the RHS of Eq.~(\ref{int 1-3/2}) and perform one more time
integration by parts and find
\beq
\nonumber
\label{int lambda>5/4}
& &\int a^{5-4\lambda}(c_{1}-\gamma a^{4(1-\lambda)})^{(3-2\lambda)/(2(\lambda-1))}da
=\dfrac{a^{2(3-2\lambda)}}{2(3-2\lambda)}(c_{1}-\gamma a^{4(1-\lambda)})^{(3-2\lambda)/(2(\lambda-1))}-\\
&-&\gamma\int a^{9-8\lambda}(c_{1}-\gamma a^{4(1-\lambda)})^{(5-4\lambda)/(2(\lambda-1))}da.
\eeq
We then express the integral on the RHS of the above equation in terms of
a hypergeometric function which leads us to the following solution valid
for $5/4<\lambda<3/2$,
\beq
\label{sol_5/4_3/2}
\nonumber
\pm Y+c_{2}&=&\dfrac{\sqrt{6}}{\kappa_{5}}\left(\dfrac{a^{2}}{2}(c_{1}-\gamma a^{4(1-\lambda)})^{1/(2(\lambda-1))}\right.\\
\nonumber
&-&\left.\dfrac{\gamma}{2(3-2\lambda)}a^{2(3-2\lambda)}(c_{1}-\gamma a^{4(1-\lambda)})^{(3-2\lambda)/(2(\lambda-1))}\right.\\ \nonumber
&+&\left.\dfrac{\gamma^{2}c_{1}^{(5-4\lambda)/(2(\lambda-1))}}{2(5-4\lambda)}a^{2(5-4\lambda)}\times\right.\\
& &\left. \times_{2}F_{1}\left(\dfrac{5-4\lambda}{2(1-\lambda)},\dfrac{5-4\lambda}{2(1-\lambda)},\dfrac{5-4\lambda}{2(1-\lambda)}+1;\dfrac{\gamma}{c_{1}}a^{4(1-\lambda)}\right)\right).
\eeq
We note again that for $\lambda<5/4$ the parameters of the hypergeometric function are
negative and this means that we have to go one more step further and
perform integration by parts on the integral on the RHS of Eq.~(\ref{int lambda>5/4}).
Following the same line of thinking, we see that to get a solution we have to refine the
values of $\lambda$ in the interval $7/6<\lambda<5/4$.
The solution we find reads
\beq
\label{sol_7/6_5/4}
\nonumber
\pm Y+c_{2}&=&\dfrac{\sqrt{6}}{\kappa_{5}}\left(\dfrac{a^{2}}{2}(c_{1}-\gamma a^{4(1-\lambda)})^{1/(2(\lambda-1))}\right.\\
\nonumber
&-&\left.\dfrac{\gamma}{2(3-2\lambda)}a^{2(3-2\lambda)}(c_{1}-\gamma a^{4(1-\lambda)})^{(3-2\lambda)/(2(\lambda-1))}\right.\\ \nonumber
&+&\left.\dfrac{\gamma^{2}}{2(5-4\lambda)}a^{2(5-4\lambda)}(c_{1}-\gamma a^{4(1-\lambda)})^{(5-4\lambda)/(2(\lambda-1))}-\right.\\ \nonumber
&-&\left.\dfrac{\gamma^{3}c_{1}^{(7-6\lambda)/(2(\lambda-1))}}{2(7-6\lambda)}a^{2(7-6\lambda)}\times\right.\\
& &\left.\times _{2}F_{1}\left(\dfrac{7-6\lambda}{2(1-\lambda)},\dfrac{7-6\lambda}{2(1-\lambda)},\dfrac{7-6\lambda}{2(1-\lambda)}+1;\dfrac{\gamma}{c_{1}}a^{4(1-\lambda)}\right)\right).
\eeq
In a similar way, we can generalize the previous results  for
$(n+1)/n<\lambda<(n-1)/(n-2)$, with $n=2k$ and $k$ a positive integer, and perform integration by parts $n/2$
times. The solution then takes the form,
\beq
\nonumber
\pm Y+c_{2}&=&\dfrac{\sqrt{6}}{\kappa_{5}}\left(\dfrac{a^{2}}{2}(c_{1}-\gamma a^{4(1-\lambda)})^{1/(2(\lambda-1))}\right.\\
\nonumber
&-&\left.\dfrac{\gamma}{2(3-2\lambda)}a^{2(3-2\lambda)}(c_{1}-\gamma a^{4(1-\lambda)})^{(3-2\lambda)/(2(\lambda-1))}\right.\\ \nonumber
&+&\left.\dfrac{\gamma^{2}}{2(5-4\lambda)}a^{2(5-4\lambda)}(c_{1}-\gamma a^{4(1-\lambda)})^{(5-4\lambda)/(2(\lambda-1))}+\ldots\right.\\ \nonumber
&+&\left.\dfrac{(-\gamma)^{n/2-1}}{2((n-1)-(n-2)\lambda)}a^{2((n-1)-(n-2)\lambda)}(c_{1}-\gamma a^{4(1-\lambda)})^{((n-1)-(n-2)\lambda)/(2(\lambda-1))}\right.\\ \nonumber
&+&\left.\dfrac{(-\gamma)^{n/2}c_{1}^{((n+1)-n\lambda)/(2(\lambda-1))}}{2((n+1)-n\lambda)}a^{2((n+1)-n\lambda)}\times \right.\\
& &\left._{2}F_{1}\left(\dfrac{(n+1)-n\lambda}{2(1-\lambda)},\dfrac{(n+1)-n\lambda}{2(1-\lambda)},\dfrac{(n+1)-n\lambda}{2(1-\lambda)}+1;\dfrac{\gamma}{c_{1}}a^{4(1-\lambda)}\right)\right).
\eeq
Direct substitution of the above equation in the field equations (\ref{syst2i})-(\ref{syst2ii})
verifies that this is indeed the general form of solution for $(n+1)/n<\lambda<(n-1)/(n-2)$.
For convenience, we can write the above equation as follows
\beq
\label{sol_n+1/n_n-1/n-2}
\nonumber
\pm Y+c_{2}&=&\frac{\sqrt{6}}{\kappa_{5}}\left(\sum_{s=0}^{n/2-1}\dfrac{(-\gamma)^{s}}{2(1-2s(\lambda-1))}\left(c_{1}a^{4(\lambda-1)}-\gamma\right)^{1/(2(\lambda-1))-s}\right.\\ \nonumber
&+&\left.\dfrac{(-\gamma)^{n/2}(c_{1})^{((n+1)-n\lambda)/(2(\lambda-1))}}{2((n+1)-n\lambda)}a^{2((n+1)-n\lambda)}\times \right.\\
&\times&\left._{2}F_{1}\left(\dfrac{(n+1)-n\lambda}{2(1-\lambda)},\dfrac{(n+1)-n\lambda}{2(1-\lambda)},\dfrac{(n+1)-n\lambda}{2(1-\lambda)}+1;\dfrac{\gamma}{c_{1}}a^{4(1-\lambda)}\right)\right).
\eeq
We next show that $\lambda>1$ is exactly the appropriate
range that provides us with regular solutions free from finite-distance
singularities. Based on this type of solution, one can construct a regular braneworld
model with matter that satisfies the strong and null energy conditions and which also successfully
localizes gravity on the flat brane.
\subsection{Regularity}
In this Subsection we study the behaviors of solutions for $\lambda>1$ and prove that they are free of finite $Y$-distance singularities.
As we saw previously, the analysis for this range of $\lambda$ leads
 to three cases, each one defined by a refinement in the
range of $\lambda$ and an implicit form of solution for the warp
factor $a(Y)$:
\begin{itemize}
\item [I.]
For $\lambda=(n+1)/n$, $n=2k$ and $k$ a positive integer,
$a(Y)$ is given by Eq.~(\ref{sol_endpoints}).

\item [II.]
For $\lambda>3/2$, $a(Y)$ is given by Eq.~(\ref{sol_3/2 above}).

\item [III.]
For $1<\lambda<3/2$ and excepting values of $\lambda$ that fall in case I,
the solution for $a(Y)$ is given by Eq.~(\ref{sol_n+1/n_n-1/n-2}) and is valid inside open
subintervals of the form $((n+1)/n,(n-1)/(n-2))$ with $n=2k$ and $k$ a positive integer such that $k\geq2$.
\end{itemize}

We begin our study with the first and simplest case I.
If we let $a\rightarrow 0^{+}$ in (\ref{sol_endpoints}), then
the RHS will approach infinity as a result of the logarithmic term that
appears there, which in turn implies that $Y\rightarrow\pm\infty$.
This means that the collapse singularity that we encountered
previously for $\lambda<1$, shifts, in this new case, to infinite distance.

For the asymptotic behaviors of $\rho$ and $p$ for $Y\rightarrow\pm\infty$,
we can deduce from Eq.~(\ref{rho to a}) and Eq.~(\ref{eos}), that they are $\rho\rightarrow (-\gamma)^{1/(1-\lambda)}$
and $p\rightarrow -(-\gamma)^{1/(1-\lambda)}$.

On the other hand, if we let $a\rightarrow\infty$, we see that all terms on the RHS of (\ref{sol_endpoints})
diverge, leading again to $Y\rightarrow\pm\infty$, while as it follows from Eqs.~(\ref{rho to a}) and (\ref{eos}),
the density and pressure vanish asymptotically. The solution described by case $I)$ is therefore
free from finite-distance singularities.

We continue with the study of case II. We note that since in Eq.~(\ref{sol_3/2 above}) the power of the argument of the
hypergeometric function is, in this case, negative, the most straightforward asymptotic
behavior implied by Eq.~(\ref{sol_3/2 above}), is that of $a\rightarrow\infty$ that makes the
argument approach zero thus leading to a convergent hypergeometric series.
Allowing $a\rightarrow\infty$, results in the divergence of the first term
on the RHS of Eq.~(\ref{sol_3/2 above})
and leads to $Y\rightarrow\pm \infty$.
The density and pressure, on the other hand, vanish asymptotically.

Having escaped the emergence of a
finite-distance singularity with a divergent warp factor,
we proceed by researching the possibility of a collapse finite-distance singularity characterized by
$a\rightarrow 0^{+}$. As was done earlier, we have to first transform the hypergeometric
function of Eq.~(\ref{sol_3/2 above}) using (\ref{erdelyi}) and then substitute  back in
Eq.~(\ref{sol_3/2 above}) to find the implicit form of solution for the warp factor. We end up with the following result,
\beq
\nonumber
\pm Y+c_{2}&=&\dfrac{\sqrt{6}}{\kappa_{5}}\left(\dfrac{a^{2}}{2}(c_{1}-\gamma a^{4(1-\lambda)})^{1/(2(\lambda-1))}+\right.\\ \nonumber
&+&\left.\dfrac{(-\gamma)^{1/(2(\lambda-1))}}{4(\lambda-1)}\sum_{n=1}^{\infty}\dfrac{\Gamma((3-2\lambda)/(2(1-\lambda))+n)}{\Gamma((3-2\lambda)/(2(1-\lambda)))n(n!)}\left(\dfrac{\gamma}{c_{1}}a^{4(1-\lambda)}\right)^{-n}+ \right.\\
&+&\left.\dfrac{(-\gamma)^{1/(2(\lambda-1))}}{2(3-2\lambda)}\left(\ln\left(-\dfrac{\gamma}{c_{1}}a^{4(1-\lambda)}\right)+\psi(1)-\psi\left(\dfrac{3-2\lambda}{2(1-\lambda)}\right)\right)\right).
\eeq
Putting $a\rightarrow 0^{+}$ in the above equation, gives $Y\rightarrow \pm\infty$ which means that the collapse type of singularity is
located at infinite distance. Also, from Eqs.~(\ref{rho to a}) and (\ref{eos}), it follows that this behavior occurs with
$\rho\rightarrow (-\gamma)^{1/(1-\lambda)}$ and $p\rightarrow -(-\gamma)^{1/(1-\lambda)}$.

Finally, the remaining case III is very similar to case II with the transformation of
Eq.~(\ref{sol_n+1/n_n-1/n-2}) according to Eq.~(\ref{erdelyi}) being in this case,
\beq
\label{erdelyi for interval}
\nonumber
&\pm& Y+c_{2}=\frac{\sqrt{6}}{2\kappa_{5}}\left(\sum_{s=0}^{n/2-1}\dfrac{(-\gamma)^{s}}{1-2s(\lambda-1)}
\left(c_{1}a^{4(\lambda-1)}-\gamma\right)^{1/(2(\lambda-1))-s}+\right.\\ \nonumber
&+&\left.(-\gamma)^{1/(2(\lambda-1))}\left(\frac{1}{2(\lambda-1)}\sum_{n=1}^{\infty}
\dfrac{\Gamma(((n+1)-n\lambda)/(2(1-\lambda))+n)}{\Gamma(((n+1)-n\lambda)/(2(1-\lambda)))n(n!)}
\left(\dfrac{\gamma}{c_{1}}a^{4(1-\lambda)}\right)^{-n}\right.\right. \\
&+&\left.\left.\dfrac{1}{(n+1)-n\lambda}\left(\ln\left(-\dfrac{\gamma}{c_{1}}a^{4(1-\lambda)}\right)
+\psi(1)-\psi\left(\dfrac{(n+1)-n\lambda}{2(1-\lambda)}\right)\right)\right)\right).
\eeq
The procedure of deriving the asymptotic behaviors implied by Eq.~(\ref{sol_n+1/n_n-1/n-2}) follows along the same steps
we performed for case II, and we encounter again the same asymptotic behaviors, namely,
\beq
\label{a_div}
a&\rightarrow&\infty \quad \textrm{and}\quad\rho\rightarrow  0,\, p\rightarrow  0, \quad \textrm{as}\quad Y\rightarrow\pm\infty\\
\label{a_0}
a&\rightarrow&0^{+} \quad \textrm{and}\quad \rho\rightarrow(-\gamma)^{1/(1-\lambda)},\, p\rightarrow -(-\gamma)^{1/(1-\lambda)},\quad \textrm{as}\quad Y\rightarrow\pm\infty.
\eeq

We have therefore shown that all three cases lead  to $\lambda$-solutions, with $\lambda>1$  that are regular in the sense
of being free from finite-distance singularities. In Sections 6, 7, we shall further explore other characteristics
of these solutions, such as the issue of localization of gravity on the brane and the matching conditions,
that make these solutions viable in the sense discussed in the Introduction.

However, before dealing with these more complicated issues in the general case,
we shall present in the next Section a worked example for the simplest possible among all the solutions in the 1-parameter
family of $\lambda$-solutions, namely the solution for $\lambda=3/2$. We shall show
that this solution shares the same basic features with all other solutions in the family,
and hence may offer clear insight into the generic behaviour of the whole set.
\section{Matching and localization: A special case}
In this Section, we focus on the solution for $\lambda=3/2$ given
in implicit form by Eq.~(\ref{sol_3/2}). Looking at Eq.~(\ref{sol_3/2}),
it is straightforward to see that it leads to the asymptotic behaviors
described by Eqs.~(\ref{a_div})-(\ref{a_0}), once we input $\lambda=3/2$
when $\lambda$ appears there. Our plan is to combine the two branches of solutions of Eq.~(\ref{sol_3/2})
and construct a matching solution that leads to a finite
Planck mass, thus rectifying the problem of localization of gravity
on the brane we had with previous models in \cite{ack3} and \cite{ack5}.

\subsection{Derivatives}
We start by writing individually the two branches of solutions of
Eq.~(\ref{sol_3/2}), these are:
\beq
\label{pos_branch}
Y^{+}&=&h_{+}(a)=\dfrac{\sqrt{6}}{\kappa_{5}}\left(\dfrac{c_{1}^{+}}{2}a^{2}-\gamma\ln a\right)+C_{2}^{+}\\
\label{neg_branch}
Y^{-}&=&h_{-}(a)=\dfrac{\sqrt{6}}{\kappa_{5}}\left(-\dfrac{c_{1}^{-}}{2}a^{2}+\gamma\ln a\right)+C_{2}^{-},
\eeq
where $Y^{\pm}$ denote the solutions we find for $Y$ using the $(\pm)$ sign in Eq.~(\ref{sol_3/2}) and
are expressed for convenience as functions, $h_{\pm}(a)$, of the warp factor, while
$c_{1}^{\pm}$, $C_{2}^{\pm}$ are the values of $c_{1}$ and $\mp c_{2}$, respectively, on the $(\pm)$ branch of $Y$.
Taking the first and second derivative of $h_{+}(a)$ we find
\beq
\label{h+'}
h_{+}'(a)&=&\dfrac{\sqrt{6}}{\kappa_{5}}\left(\dfrac{c_{1}^{+}a^{2}-\gamma}{a}\right)\\
\label{h+''}
h_{+}''(a)&=&\dfrac{\sqrt{6}}{\kappa_{5}}\dfrac{c_{1}^{+}}{a^{2}}\left(a^{2}+\dfrac{\gamma}{c_{1}^{+}}\right).
\eeq
Now, since $\gamma<0$ and $c_{1}^{+}>0$, the first derivative is strictly positive,
while the second derivative changes from negative on $(0,\sqrt{-\gamma/c_{1}^{+}})$
to positive on $(\sqrt{-\gamma/c_{1}^{+}},\infty)$ and crosses zero at $a=\sqrt{-\gamma/c_{1}^{+}}$,
making the point $(\sqrt{-\gamma/c_{1}^{+}},-\sqrt{6}/\kappa_{5}(\gamma/2+\gamma \ln \sqrt{-\gamma/c_{1}^{+}})+C_{2}^{+})$
an inflection point of the graph of $h_{+}$.

Calculating the first two derivatives of $h_{-}(a)$, on the other
hand, gives similar results the only difference being
that the signs of the derivatives are the opposites of the ones of $h_{+}$ given
in Eqs.~(\ref{h+'})-(\ref{h+''}) above. This means that $h'_{-}$ is strictly negative and
$h''_{-}$ changes from positive on $(0,\sqrt{-\gamma/c_{1}^{-}})$ to negative on $(\sqrt{-\gamma/c_{1}^{-}},\infty)$
with $(\sqrt{-\gamma/c_{1}^{-}},\sqrt{6}/\kappa_{5}(\gamma/2+\gamma \ln \sqrt{-\gamma/c_{1}^{-}})+C_{2}^{-})$
being an inflection point of the graph of $h_{-}$.

\subsection{Matching at the inflection point}
As we saw in Section 3, convergence of the power series expansion of the integral of the hypergeometric series given by Eq. (\ref{a_sing}) is effected as long as $z$ is not a real number bigger than unity. 

This means that for $\lambda=3/2$, we must have,
\be
a>\left|\frac{\gamma}{c_1}\right|^{1/2},
\ee
but as we have already remarked, convergence may be extended by analytic continuation to regions far beyond that, where the opposite inequality will be true. The important thing is that because of this, we can match the two branches at the inflection point of their graphs found above, because the convergence of the matching solution is only consistent in that case.

A natural assumption is that the warp factor should be continuous there so
\be
\label{match cond1}
\sqrt{\dfrac{-\gamma}{c_{1}^{+}}}=\sqrt{\dfrac{-\gamma}{c_{1}^{-}}},\quad \textrm{or}
,\quad c_{1}^{+}=c_{1}^{-}=c_{1}.
\ee
Also, the inflection point should have to the same $Y$ coordinate through the two branches
which means that
\be
h_{+}\left(\sqrt{\dfrac{-\gamma}{c_{1}}}\right)=h_{-}\left(\sqrt{\dfrac{-\gamma}{c_{1}}}\right)
\ee
and yields the following relation between $C_{2}^{+}$ and $C_{2}^{-}$,
\be
\label{match cond2}
-\dfrac{\sqrt{6}}{\kappa_{5}}\left(\dfrac{\gamma}{2}+\gamma \ln \sqrt{\dfrac{-\gamma}{c_{1}}}\right)+C_{2}^{+}=\dfrac{\sqrt{6}}{\kappa_{5}}\left(\dfrac{\gamma}{2}+\gamma \ln \sqrt{\dfrac{-\gamma}{c_{1}}}\right)+C_{2}^{-},
\ee
or,
\be
C_{2}^{+}=\dfrac{\sqrt{6}}{\kappa_{5}}\left(\gamma+2\gamma\ln \sqrt{\dfrac{-\gamma}{c_{1}}}\right)+C_{2}^{-}.
\ee
The matching graphs that we obtain in this way, have an axis of symmetry at $Y=Y_{s}$, with $Y_{s}$ given by
\be
\label{Y_s}
Y_{s}=h_{\pm}(\sqrt{-\gamma/c_{1}})=-\dfrac{\sqrt{6}}{\kappa_{5}}\left(\dfrac{\gamma}{2}+\gamma \ln \sqrt{\dfrac{-\gamma}{c_{1}}}\right)+C_{2}^{+}=\dfrac{\sqrt{6}}{\kappa_{5}}\left(\dfrac{\gamma}{2}+\gamma \ln \sqrt{\dfrac{-\gamma}{c_{1}}}\right)+C_{2}^{-}
\ee
and designating the position of the brane. For simplicity we can place the brane at
$Y=0$ by putting $Y_{s}=0$ in $(\ref{Y_s})$ and calculating the values of
$C_{2}^{\pm}$ that make this possible. We find
\be
\label{C_2_3/2}
C_{2}^{+}=\dfrac{\sqrt{6}}{\kappa_{5}}\left(\dfrac{\gamma}{2}+\gamma \ln \sqrt{\dfrac{-\gamma}{c_{1}}}\right)=-C_{2}^{-}.
\ee
The matching solution described by Eq. (\ref{pos_branch}) and (\ref{neg_branch}) that satisfies the above boundary condition
can be written as
\be
\label{matching sol 3/2}
|Y|=\dfrac{\sqrt{6}}{\kappa_{5}}\left(-\dfrac{c_{1}}{2}a^{2}+\gamma\ln a-\dfrac{\gamma}{2}-\gamma \ln \sqrt{\dfrac{-\gamma}{c_{1}}}\right),\quad 0<a\leq\sqrt{\dfrac{-\gamma}{c_{1}}}.
\ee
We note that $\rho$ is well defined and continuous at the location of the brane
$Y=0$ and its value is
\be
\rho\left(\sqrt{\dfrac{-\gamma}{c_{1}}}\right)\equiv\rho(0)=\dfrac{1}{4\gamma^{2}},
\ee
where $\rho(0)$ denotes $\rho(Y=0)$ and it is an abbreviation of
$\rho(h_{\pm}^{-1}(\sqrt{-\gamma/c_{1}}))$.

In Figs.1, 2 and 3, we depict the graphs of $h_{+}$, $h_{-}$ and the matching branches, respectively.
In Fig.~4, we have rotated the axes of Fig.~3, to gain a more convenient view of the evolution of the
warp factor as a function of $Y$. The reason for constructing the matching solution by keeping the solid lines
instead of the dashed ones in Fig.~3, is to obtain a finite Planck mass, as we show in the analysis below.

\begin{center}
\includegraphics[scale=1.0]{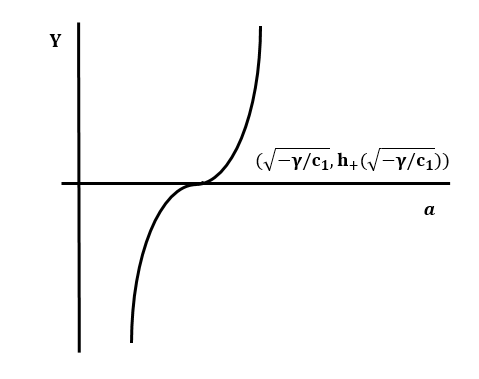}
\captionof{figure}{Graph of $h_{+}(a)$.}
\end{center}

\begin{center}
\includegraphics[scale=1.0]{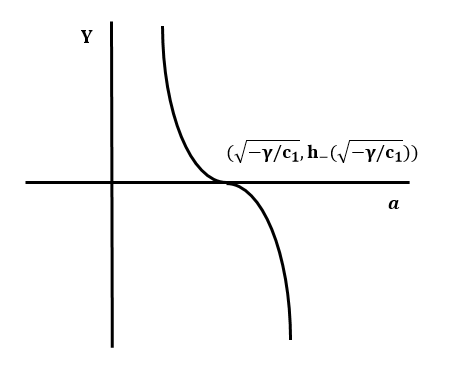}
\captionof{figure}{Graph of $h_{-}(a)$.}
\end{center}

\begin{center}
\includegraphics[scale=1.0]{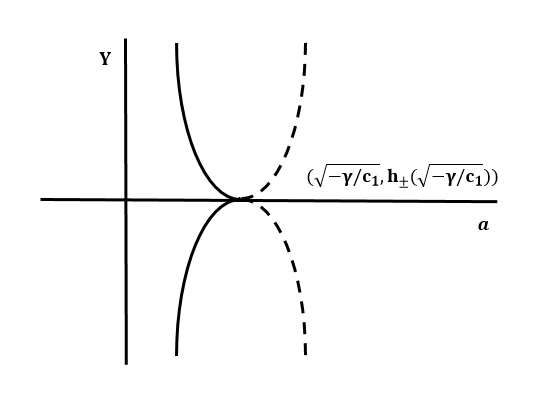}
\captionof{figure}{Matching graphs at the inflection point.}
\end{center}

\begin{center}
\includegraphics[scale=1.0]{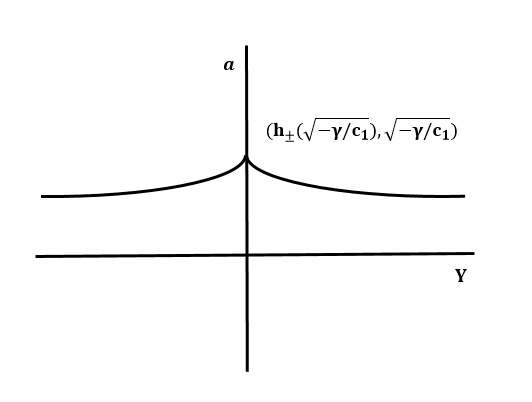}
\captionof{figure}{Graph of the warp factor as a function of $Y$, for $\lambda=3/2$.}
\end{center}

Next, we take into account the jump of the derivative of
the warp factor across the brane and find, for our type of geometry, the following
junction condition
\be
\label{match_3}
a'(0^{+})-a'(0^{-})=2a'(0^{+}) =-\kappa_{5}^{2}\dfrac{f(\rho(0))a(0)}{3},
\ee
where $f(\rho(0))$ is the tension of the brane. Using Eq.~(\ref{syst2i}),
in the above equation we can derive the relation between the brane tension
and the density, this reads
\be
\label{match_4}
f(\rho(0))=-\dfrac{\sqrt{6\rho(0)}}{\kappa_{5}}=\dfrac{\sqrt{6}}{2\kappa_{5}\gamma}.
\ee

\subsection{Localization}
Finally, we calculate the Planck mass for the matching solution.
The value of the four-dimensional Planck mass, $M_{p}^{2}=8\pi/\kappa$,
is determined by the following integral (see \cite{ack3}, \cite{ack5} and \cite{forste}),
\be
\label{mass}
\frac{\kappa_{5}^{2}}{\kappa}=\int_{-Y_{c}}^{Y_{c}}a^{2}(Y)dY.
\ee
For the solution of Eq.~(\ref{matching sol 3/2}), the behaviour of $a^{2}$ as $Y\rightarrow-\infty$ is
\be
\label{sq a sol 3/2}
a^{2}\sim e^{-(\sqrt{6}\kappa_{5}/(3\gamma))Y},
\ee
and using the symmetry of Eq.~(\ref{matching sol 3/2}), we can write the above integral in the following form
\be
\int_{-Y_{c}}^{Y_{c}}a^{2}dY=2\int_{-Y_{c}}^{0}a^{2}dY\sim
2\int_{-Y_{c}}^{0}e^{-(\sqrt{6}\kappa_{5}/(3\gamma))Y}dY=
-\sqrt{6}\dfrac{\gamma}{\kappa_{5}} (1-e^{(\sqrt{6}\kappa_{5}/(3\gamma))Y_{c}}).
\ee
In the limit $Y_{c}\rightarrow\infty$ and since $\gamma<0$, we see that the Planck mass remains
finite and is proportional to $$-\sqrt{6}\dfrac{\gamma}{\kappa_{5}}.$$

Summarizing the case of study of $\lambda=3/2$, we have constructed a
regular matching solution that satisfies both the null and strong energy
conditions and at the same time successfully localizes gravity on the brane.
\section{Matching: The general case}
Now, we generalize the matching mechanism we constructed
in the previous Section for the case of $\lambda=3/2$, to cover the broader case
of $\lambda>1$. As mentioned before, the two cases share many
common features and we can use the example of $\lambda=3/2$ as a
guide to the most complicated case of $\lambda>1$.

We begin by writing separately the two solutions we get for
$\lambda=(n+1)/n$, $n=2k$, k a positive integer given
by Eq.~(\ref{sol_endpoints}), we have
\be
\label{sol_endpoints_pos_branch}
Y^{+}=h_{+}(a)=\frac{\sqrt{6}}{\kappa_{5}}\left(\sum_{s=0}^{k-1}\dfrac{k!}{(k-s)!s!}\dfrac{(c_{1}^{+})^{k-s}}{2-2s/k} a^{2-2s/k}(-\gamma)^{s}+(-\gamma)^{k}\ln a\right)+C_{2}^{+}
\ee
\be
\label{sol_endpoints_neg_branch}
Y^{-}=h_{-}(a)=-\frac{\sqrt{6}}{\kappa_{5}}\left(\sum_{s=0}^{k-1}\frac{k!}{(k-s)!s!}\frac{(c_{1}^{-})^{k-s}}{2-2s/k} a^{2-2s/k}(-\gamma)^{s}+(-\gamma)^{k}\ln a\right)+C_{2}^{-},
\ee
where the notation of $Y^{\pm}$, $c_{1}^{\pm}$ and $C_{2}^{\pm}$ represents, as in Section 5, the values of the
variables/constants of the $(\pm)$ choice of solution respectively.

To evaluate easily the first two derivatives of $h_{+}$, we note from
Eq.~(\ref{hyper_integral}) that $h'_{+}$ is equal to the integrand
function that appears there. After some manipulation we can rewrite
the integrand function as $(c_{1}^{+}a^{1/k}-\gamma a^{-1/k})^{k}$ and
then differentiate once more to evaluate $h''_{+}$, we obtain
\beq
h'_{+}(a)&=&\frac{\sqrt{6}}{\kappa_{5}}(c_{1}^{+}a^{1/k}-\gamma a^{-1/k})^{k}\\
h''_{+}(a)&=&\frac{\sqrt{6}}{\kappa_{5}}a^{-1/k-1}(c_{1}^{+}a^{1/k}-\gamma a^{-1/k})^{k-1}(c_{1}^{+}a^{2/k}+\gamma).
\eeq
Since $\gamma<0$ and $c_{1}^{+}>0$, $h'_{+}$ is always positive
and $h''_{+}$ changes from negative on $(0,(-\gamma/c_{1}^{+})^{k/2})$ to positive
on $((-\gamma/c_{1}^{+})^{k/2},\infty)$, so that the point
$$((-\gamma/c_{1}^{+})^{k/2},h_{+}((-\gamma/c_{1}^{+})^{k/2}))$$
is an inflection point of the graph of $h_{+}$. As in Section 5, we can deduce the
behavior of the first two derivatives of $h_{-}$ from those of $h_{+}$ since they are
opposites.

As in the previous Section, we are allowed to join the two solutions of
Eqs.~(\ref{sol_endpoints_pos_branch})-(\ref{sol_endpoints_neg_branch})
at their common inflection point.  This implies that we take,
\be
c_{1}^{+}=c_{1}^{-}=c_{1},
\ee
and
\be
h_{+}\left(\left(\dfrac{-\gamma}{c_{1}}\right)^{k/2}\right)=h_{-}\left(\left(\dfrac{-\gamma}{c_{1}}\right)^{k/2}\right).
\ee
The latter equation reveals the relation between the two constants $C_{2}^{\pm}$
\be
C_{2}^{+}=-\frac{2\sqrt{6}}{\kappa_{5}}\left(\sum_{s=0}^{k-1}\dfrac{k!}{(k-s)!s!}\dfrac{(-\gamma)^{k}}{2-2s/k}+(-\gamma)^{k}\ln \left(\dfrac{-\gamma}{c_{1}}\right)^{k/2}\right)+C_{2}^{-}.
\ee
The matching graphs have an axis of symmetry at $Y=Y_{s}$, with
\beq
\nonumber
\label{Y_s_endpoints}
Y_{s}&=&h_{\pm}((-\gamma/c_{1})^{k/2}))=\frac{\sqrt{6}}{\kappa_{5}}\left(\sum_{s=0}^{k-1}\dfrac{k!}{(k-s)!s!}\dfrac{(-\gamma)^{k}}{2-2s/k}+(-\gamma)^{k}\ln \left(\dfrac{-\gamma}{c_{1}}\right)^{k/2}\right)+C_{2}^{+}\\
&=&-\frac{\sqrt{6}}{\kappa_{5}}\left(\sum_{s=0}^{k-1}\dfrac{k!}{(k-s)!s!}\dfrac{(-\gamma)^{k}}{2-2s/k}+(-\gamma)^{k}\ln \left(\dfrac{-\gamma}{c_{1}}\right)^{k/2}\right)+C_{2}^{-},
\eeq
hence $Y_{s}$ is where we can position the brane. To simplify our expressions
we take again $Y_{s}=0$, which means that the arbitrary constants $C_{2}^{\pm}$
are set to the values
\be
\label{C_2 endpoints}
C_{2}^{+}=-\frac{\sqrt{6}}{\kappa_{5}}\left(\sum_{s=0}^{k-1}\dfrac{k!}{(k-s)!s!}\dfrac{(-\gamma)^{k}}{2-2s/k}+(-\gamma)^{k}\ln \left(\dfrac{-\gamma}{c_{1}}\right)^{k/2}\right)=-C_{2}^{-}.
\ee
The matching solution described by Eqs.~(\ref{sol_endpoints_pos_branch}) and
(\ref{sol_endpoints_neg_branch}) that satisfies the boundary condition
(\ref{C_2 endpoints}) 
then takes the form
\beq
\nonumber
\label{match_endpoints}
|Y|&=&\frac{\sqrt{6}}{\kappa_{5}}
\left(\sum_{s=0}^{k-1}\frac{k!}{(k-s)!s!(2-2s/k)}\left(-c_{1}^{k-s}a^{2-2s/k}(-\gamma)^{s}+(-\gamma)^{k}\right)+\right.\\
&+&\left.(-\gamma)^{k}\left(\ln \left(\dfrac{-\gamma}{c_{1}}\right)^{k/2}-\ln a\right)\right), \quad 0<a\leq\left(\dfrac{-\gamma}{c_{1}}\right)^{k/2}
\eeq
and is depicted in Fig.~5.
\begin{center}
\includegraphics[scale=1.0]{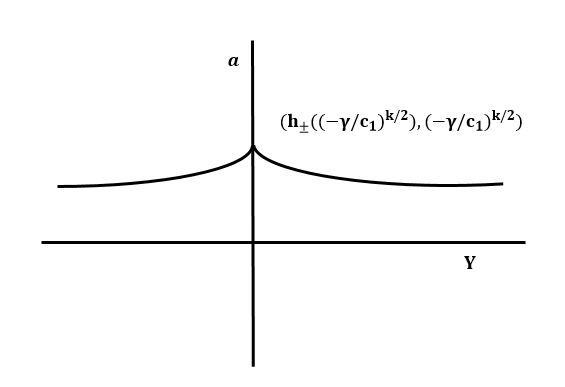}
\captionof{figure}{Graph of the warp factor as a function of $Y$, for $\lambda=(2k+1)/(2k)$.}
\end{center}

Again $\rho$ is well defined and continuous at the location of the brane and is equal to
\be
\rho\left(\left(\dfrac{-\gamma}{c_{1}}\right)^{k/2}\right)\equiv\rho(0)=\dfrac{1}{(-2\gamma)^{2k}}.
\ee
Also, the junction condition for the jump of $a'$, is identical with the case of the
previous Section and it is therefore described by Eq.~(\ref{match_3}), while the
form of the brane tension that follows from (\ref{match_3}) 
is now
\be
f(\rho(0))=-\dfrac{\sqrt{6\rho(0)}}{\kappa_{5}}=-\dfrac{\sqrt{6}}{\kappa_{5}(-2\gamma)^{k}}.
\ee

We repeat the procedure for all other values of $1<\lambda<3/2$ that lie on the
intervals $((n+1)/n,(n-1)/(n-2))$. The two solutions described by
Eq.~(\ref{sol_n+1/n_n-1/n-2}) are
\beq
\label{sol_n+1/n_n-1/n-2_pos_branch}
\nonumber
Y^{+}&=&h_{+}(a)=\frac{\sqrt{6}}{\kappa_{5}}\left(\sum_{s=0}^{n/2-1}\dfrac{(-\gamma)^{s}}{2(1-2s(\lambda-1))}\left(c_{1}^{+}a^{4(\lambda-1)}-\gamma\right)^{1/(2(\lambda-1))-s}\right.\\ \nonumber
&+&\left.\dfrac{(-\gamma)^{n/2}(c_{1}^{+})^{((n+1)-n\lambda)/(2(\lambda-1))}}{2((n+1)-n\lambda)}a^{2((n+1)-n\lambda)}\times \right.\\
&\times&\left._{2}F_{1}\left(\dfrac{(n+1)-n\lambda}{2(1-\lambda)},
\dfrac{(n+1)-n\lambda}{2(1-\lambda)},\dfrac{(n+1)-n\lambda}{2(1-\lambda)}+1;\dfrac{\gamma}
{c_{1}^{+}}a^{4(1-\lambda)}\right)\right)\nonumber\\
&+&C_{2}^{+}
\eeq
and
\beq
\nonumber
\label{sol_n+1/n_n-1/n-2_neg_branch}
Y^{-}&=&h_{-}(a)=-\frac{\sqrt{6}}{\kappa_{5}}\left(\sum_{s=0}^{n/2-1}\dfrac{(-\gamma)^{s}}{2(1-2s(\lambda-1))}\left(c_{1}^{-}a^{4(\lambda-1)}-\gamma\right)^{1/(2(\lambda-1))-s} \right.\\ \nonumber
&+&\left.\dfrac{(-\gamma)^{n/2}(c_{1}^{-})^{((n+1)-n\lambda)/(2(\lambda-1))}}{2((n+1)-n\lambda)}a^{2((n+1)-n\lambda)}\times \right.\\
&\times&\left._{2}F_{1}\left(\dfrac{(n+1)-n\lambda}{2(1-\lambda)},\dfrac{(n+1)-n\lambda}{2(1-\lambda)},
\dfrac{(n+1)-n\lambda}{2(1-\lambda)}+1;\dfrac{\gamma}{c_{1}^{-}}a^{4(1-\lambda)}\right)\right)\nonumber\\
&+&C_{2}^{-}.
\eeq
By differentiating Eq.~(\ref{sol_n+1/n_n-1/n-2_pos_branch}) with respect to $a$,
we arrive simply at the integrand function on the left-hand side of Eq.~(\ref{hyper_integral}).
After that, finding $h''_{+}$ is straightforward
and the first two derivatives of $h_{+}$ read
\beq
h'_{+}(a)&=&\frac{\sqrt{6}}{\kappa_{5}}\dfrac{a}{(c_{1}^{+}-\gamma a^{4(1-\lambda)})^{1/2(1-\lambda)}}=\frac{\sqrt{6}}{\kappa_{5}}(c_{1}^{+}a^{2(\lambda-1)}-\gamma a^{2(1-\lambda)})^{1/(2(\lambda-1))}\\
h''_{+}(a)&=&\frac{\sqrt{6}}{\kappa_{5}}a^{2\lambda-3}(c_{1}^{+}+\gamma a^{4(1-\lambda)})
(c_{1}^{+}a^{2(\lambda-1)}-\gamma a^{2(1-\lambda)})^{(3-2\lambda)/(2(\lambda-1))}.
\eeq
For our choice of parameters, $h'_{+}$ is always positive
and $h''_{+}$ changes from negative on $(0,(-\gamma/c_{1}^{+})^{1/(4(\lambda-1))})$ to positive
on $((-\gamma/c_{1}^{+})^{1/(4(\lambda-1))},\infty)$, thus the point
$$((-\gamma/c_{1}^{+})^{1/(4(\lambda-1))},h_{+}((-\gamma/c_{1}^{+})^{1/(4(\lambda-1))}))$$
is an inflection point of the graph of $h_{+}$. As before, the
behavior of the first two derivatives of $h_{-}$ can be obtained from
those of $h_{+}$ since they are opposites.

Joining the two solutions of
Eqs.~(\ref{sol_n+1/n_n-1/n-2_pos_branch})-(\ref{sol_n+1/n_n-1/n-2_neg_branch})
at their common inflection point implies
\be
c_{1}^{+}=c_{1}^{-}=c_{1}
\ee
and
\be
h_{+}\left(\left(\dfrac{-\gamma}{c_{1}}\right)^{1/(4(\lambda-1))}\right)=h_{-}\left(\left(\dfrac{-\gamma}{c_{1}}\right)^{1/(4(\lambda-1))}\right)
\ee
and thus provides the following relation between the two constants $C_{2}^{\pm}$
\beq
\nonumber
C_{2}^{+}&=&-\frac{\sqrt{6}}{\kappa_{5}}(-\gamma)^{1/(2(\lambda-1))}\left(\sum_{s=0}^{n/2-1}\dfrac{2^{1/(2(\lambda-1))-s}}{1-2s(\lambda-1)}+\dfrac{1}{(n+1)-n\lambda}\times\right.\\
&\times&\left._{2}F_{1}\left(\dfrac{(n+1)-n\lambda}{2(1-\lambda)},\dfrac{(n+1)-n\lambda}{2(1-\lambda)},\dfrac{(n+1)-n\lambda}{2(1-\lambda)}+1;-1\right)\right)
+C_{2}^{-}.
\eeq
The axis of symmetry is located at $Y=Y_{s}$, with $Y_{s}$ being equal to
\beq
\nonumber
\label{Y_s_interval}
Y_{s}&=&h_{\pm}((-\gamma/c_{1})^{1/(4(\lambda-1))}))=\frac{\sqrt{6}}{2\kappa_{5}}(-\gamma)^{1/(2(\lambda-1))}\left(\sum_{s=0}^{n/2-1}\dfrac{2^{1/(2(\lambda-1))-s}}{1-2s(\lambda-1)}+\right. \\ \nonumber
&+&\left.\dfrac{1}{(n+1)-n\lambda}\times _{2}F_{1}\left(\dfrac{(n+1)-n\lambda}{2(1-\lambda)},\dfrac{(n+1)-n\lambda}{2(1-\lambda)},\dfrac{(n+1)-n\lambda}
{2(1-\lambda)}+1;-1\right)\right)\\ \nonumber
&+&C_{2}^{+}=\\ \nonumber
&=&-\frac{\sqrt{6}}{2\kappa_{5}}(-\gamma)^{1/(2(\lambda-1))}\left(\sum_{s=0}^{n/2-1}\dfrac{2^{1/(2(\lambda-1))-s}}{1-2s(\lambda-1)}+\dfrac{1}{((n+1)-n\lambda)}\times \right.\\\nonumber
&\times&\left._{2}F_{1}\left(\dfrac{(n+1)-n\lambda}{2(1-\lambda)},\dfrac{(n+1)-n\lambda}{2(1-\lambda)},
\dfrac{(n+1)-n\lambda}{2(1-\lambda)}+1;-1\right)\right)\\
&+&C_{2}^{-},
\eeq
and labeling the position of the brane. Positioning the brane at $Y=0$, translates
to choosing $Y_{s}=0$ which determines the values of the constants $C_{2}^{\pm}$ in the following way
\beq
\nonumber
\label{C_2_interval}
C_{2}^{+}&=&-\frac{\sqrt{6}}{2\kappa_{5}}(-\gamma)^{1/(2(\lambda-1))}\left(\sum_{s=0}^{n/2-1}\dfrac{2^{1/(2(\lambda-1))-s}}{1-2s(\lambda-1)}+\dfrac{1}{(n+1)-n\lambda}\times\right. \\
&\times&\left._{2}F_{1}\left(\dfrac{(n+1)-n\lambda}{2(1-\lambda)},\dfrac{(n+1)-n\lambda}{2(1-\lambda)},\nonumber
\dfrac{(n+1)-n\lambda}{2(1-\lambda)}+1;-1\right)\right)\\
&=&-C_{2}^{-}.
\eeq
The matching solution that satisfies the above boundary condition is given by,
\beq
\label{match interval}
|Y|&=&\frac{\sqrt{6}}{2\kappa_{5}}\left([\textrm{sum}]-\dfrac{(-\gamma)^{n/2}(c_{1})^{((n+1)-n\lambda)/(2(\lambda-1))}}{(n+1)-n\lambda}a^{2((n+1)-n\lambda)}\times
\right.\nonumber\\
&\times&\left._{2}F_{1}\left(\dfrac{(n+1)-n\lambda}{2(1-\lambda)},
\dfrac{(n+1)-n\lambda}{2(1-\lambda)},\dfrac{(n+1)-n\lambda}{2(1-\lambda)}+1;\dfrac{\gamma}{c_{1}}a^{4(1-\lambda)}\right)
+\dfrac{(-\gamma)^{1/(2(\lambda-1))}}{(n+1)-n\lambda}\times\right. \nonumber\\
&\times&\left._{2}F_{1}\left(\dfrac{(n+1)-n\lambda}{2(1-\lambda)},\dfrac{(n+1)-n\lambda}{2(1-\lambda)},
\dfrac{(n+1)-n\lambda}{2(1-\lambda)}+1;-1\right)\right),
\eeq
where
$$ 0<a\leq\left(\dfrac{-\gamma}{c_{1}}\right)^{1/4(\lambda-1)}$$
and we have set the expression $[\textrm{sum}]$ to be equal to,
$$
\sum_{s=0}^{n/2-1}\dfrac{1}{1-2s(\lambda-1)}\left((-\gamma)^{1/(2(\lambda-1))}2^{1/(2(\lambda-1))-s}
-(-\gamma)^{s}\left(c_{1}a^{4(\lambda-1)}-\gamma\right)^{1/(2(\lambda-1))-s}\right).
$$
This matching solution is then depicted in Fig.~6.

\begin{center}
\includegraphics[scale=1.0]{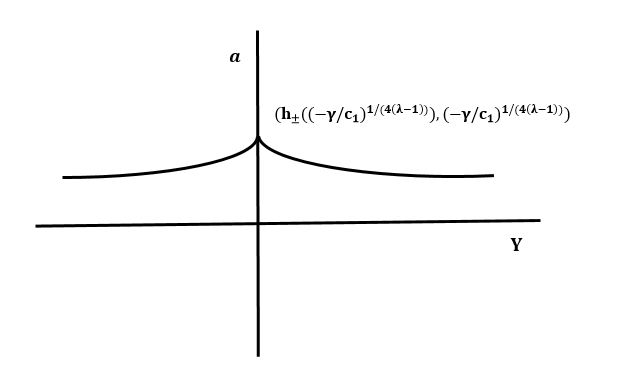}
\captionof{figure}{Graph of the warp factor as a function of $Y$, for
$(n+1)/n<\lambda<(n-1)/(n-2)$.}
\end{center}

The value of $\rho$ there is equal to
\be
\rho\left(\left(\dfrac{-\gamma}{c_{1}}\right)^{1/(4(\lambda-1))}\right)\equiv\rho(0)=(-2\gamma)^{1/(1-\lambda)}.
\ee
Once more, the junction condition for the jump of $a'$, is given by Eq.~(\ref{match_3}), whereas the brane tension becomes now
\be
f(\rho(0))=-\dfrac{\sqrt{6\rho(0)}}{\kappa_{5}}=-\dfrac{\sqrt{6}}{\kappa_{5}}(-2\gamma)^{1/(2(1-\lambda))}.
\ee
Finally, for $\lambda>3/2$, the construction of a matching solution follows along the same
steps since the solution of Eq.~(\ref{sol_3/2 above}) can be obtained from Eq.~(\ref{sol_n+1/n_n-1/n-2})
by inputting $n=2$.
\section{Localization}
We are going to calculate the Planck mass for the matching solutions
we constructed in the previous Section, for $\lambda>1$.

First, we derive the behaviour of $a^{2}$ as $Y\rightarrow-\infty$
from Eq.~(\ref{match_endpoints}),
we find
\be
\label{a_sq n+1/n}
a^{2}\sim e^{\sqrt{6}\kappa_{5}Y(-\gamma)^{-k}/3}.
\ee
We can write the integral of Eq.~(\ref{mass}) in the following form
\beq
\nonumber
\int_{-Y_{c}}^{Y_{c}}a^{2}dY&=&2\int_{-Y_{c}}^{0}a^{2}dY\sim
2\int_{-Y_{c}}^{0}e^{\sqrt{6}\kappa_{5}Y(-\gamma)^{-k}/3}dY=\\
&=&\frac{\sqrt{6}}{\kappa_{5}}(-\gamma)^{k} (1-
e^{-\sqrt{6}\kappa_{5}Y_{c}(-\gamma)^{-k}/3}).
\eeq
Taking $Y_{c}\rightarrow\infty$ and since $\gamma<0$, we find that the Planck mass remains
finite and is proportional to the value
\be
\dfrac{\sqrt{6}}{\kappa_{5}}(-\gamma)^{k}.
\ee

Finally, we deduce the behaviour of $a^{2}$ as $Y\rightarrow-\infty$
by transforming Eq.~(\ref{match interval}) according to Eq.~(\ref{erdelyi for interval}),
we obtain
\be
\label{a_sq n+1/n_n-1/n-2}
a^{2}\sim \left(-\dfrac{c_{1}}{\gamma}\right)^{1/(2(1-\lambda))}e^{\kappa_{5}((n+1)-n\lambda)(-\gamma)^{1/(2(1-\lambda))}Y/(\sqrt{6}(1-\lambda))}.
\ee
The integral of Eq.~(\ref{mass}) can then be written as
\be
\nonumber
\int_{-Y_{c}}^{Y_{c}}a^{2}dY=2\int_{-Y_{c}}^{0}a^{2}dY\sim
2\int_{-Y_{c}}^{0}\left(-\frac{c_{1}}{\gamma}\right)^{1/(2(1-\lambda))}e^{\kappa_{5}((n+1)-n\lambda)(-\gamma)^{1/(2(1-\lambda))}Y/(\sqrt{6}(1-\lambda))}dY=
\ee
\be
=\frac{2\sqrt{6}(-\gamma)^{1/(2(\lambda-1))}(1-\lambda)}{\kappa_{5}((n+1)-n\lambda)}
\left(-\frac{c_{1}}{\gamma}\right)^{1/(2(1-\lambda))}(1-e^{-\kappa_{5}((n+1)-n\lambda)(-\gamma)^{1/(2(1-\lambda))}Y_{c}/(\sqrt{6}(1-\lambda))}).
\ee
As $Y_{c}\rightarrow\infty$ and since $\gamma<0$ and $((n+1)-n\lambda)/(1-\lambda)>0$, the Planck mass remains
finite and is proportional to the value
\be
\label{mplanck interval}
\frac{2\sqrt{6}(-\gamma)^{1/(\lambda-1)}(1-\lambda)c_{1}^{1/(2(1-\lambda))}}{\kappa_{5}((n+1)-n\lambda)}.
\ee

We can check the behavior of the Planck mass for $(n+1)/n<\lambda<(n-1)/(n-2)$ and
as $\lambda$ approaches one. We can choose $\lambda=1+1/(n-1)$, then clearly
$\lambda$ lies on the interval $((n+1)/n,(n-1)/(n-2))$ and approaches one from the right
as $n\rightarrow \infty$. Inputting this value in (\ref{mplanck interval}), we find
\be
\frac{2\sqrt{6}}{\kappa_{5}}\left(-\dfrac{\gamma}{\sqrt{c_{1}}}\right)^{n-1},
\ee
from which we see that for large values of $n$, the behavior of the Planck mass depends on
the ordering between $-\gamma$ and $c_{1}$:
\be
\left(-\dfrac{\gamma}{\sqrt{c_{1}}}\right)^{n-1}\rightarrow \left\{ \begin{array}{c}
                                                                      0, -\gamma<\sqrt{c_{1}}\\
                                                                      1, -\gamma=\sqrt{c_{1}}\\
                                                                      \infty, -\gamma>\sqrt{c_{1}}.
                                                                    \end{array}
                                                                    \right.
\ee

Summarizing, all solutions for $\lambda>1$, are free from finite-distance
singularities, they are compatible with the strong and null energy conditions
and at the same time lead to a finite Planck mass without imposing
any further conditions on the parameters of our model.
All solutions for a linear equation of state $(\lambda=1$) on the other hand,
suffered from finite-distance singularities that could only be avoided by a
matching mechanism \cite{ack3}. Still as we showed in \cite{ack3}, for the
regular matching solutions, the requirement of a finite Planck mass, leads
to a further restriction on $\gamma$, namely that of $-2<\gamma<-1$,
which turns out to be incompatible with the requirements of the energy
conditions that confine $\gamma$ to values at least greater than $-1$.
We therefore see that the effect of the non-linear equation of state is significant in the sense that it
rectifies previous findings that were based on a linear equation of state.
\section{Conclusions}
In this paper, we studied the behaviour of solutions of a 5d brane-world equations in the bulk, with particular emphasis on their regularity properties and associated possibility of localisation of gravity on a flat brane. We showed that the behaviour of all solutions depends critically on the non-linear equation of state parameter $\lambda$ ($p=\gamma\rho^\lambda$). We showed that solutions with $\lambda<1$ cannot be regular and at the same time localise gravity on the brane. However, the main result of this work is that when $\lambda>1$, there are solutions which are free of singularities and having finite Planck mass, that is they are able to localise gravity on the brane. We also showed that the behaviour of the matched solutions is consistent in the sense that they are convergent in the correct intervals for their matching to be allowed. It is an interesting open question whether such non-linear equation of state satisfying the null energy condition can be realised using an underlying microscopic description.

\section*{Acknowledgments}
Work partially performed by I.A. as International professor of the Francqui Foundation, Belgium.

\end{document}